\documentclass[11pt]{article}
\usepackage[dvips]{color}
\usepackage{epsfig}
\usepackage{amsmath}
\usepackage{graphicx}
\date{}
\begin{document}
\title{{\bf Some Features of Scattering Problem in a $\kappa$-Deformed Minkowski Spacetime}}

\author{M. Khodadi$^1$\thanks{%
m.khodadi@stu.umz.ac.ir},\,\, K. Nozari$^1$\thanks{%
knozari@umz.ac.ir}
\\\\
$^1${\small {\it Department of Physics, Faculty of Basic Sciences,
University of Mazandaran,}}\\ {\small {\it P.O. Box
47416-95447, Babolsar, Iran}}}

\maketitle

\title{}
\begin{abstract}
The doubly special relativity (DSR) theories are suggested in order to incorporate
an observer-independent length scale in special theory of relativity.
The Magueijo-Smolin proposal of DSR is realizable through a
particular form of the noncommutative (NC) spacetime (known as $\kappa$-Minkowski spacetime)
in which the Lorentz symmetry is preserved. In this framework, the NC parameter $\kappa$ provides the origin
of natural cutoff energy scale. Using a nonlinear
deformed relativistic dispersion relation along with the Lorentz
transformations, we investigate some phenomenological facets of two-body
collision problem (without creation of new particles) in a $\kappa$-Minkowski spacetime.
By treating an elastic scattering problem, we study
effects of the Planck scale energy cutoff on some relativistic kinematical
properties of this scattering problem. The results are challenging in the
sense that as soon as one turns on the $\kappa$-spacetime
extension, the nature of the two-body collision alters from elastic to inelastic one.
It is shown also that a significant kinematical variable involving in heavy
ion collisions, the rapidity, is not essentially an additive quantity under
a sequence of the nonlinear representation of the Lorentz transformations.
\\
\vspace{0.8mm}\newline \textbf{Keywords}: Doubly Special Relativity; Noncommutative Spacetime;
Nonlinear Lorentz Transformations; Elastic Scattering; Rapidity.
\end{abstract}
\maketitle
\section{Introduction}
Lorentz symmetry (LS), from both theoretical and experimental perspectives, has a
very special status in modern physics so that it has attracted much attention
these years. More than half a century, we are faced with a wide variety of researches
that are focused on verifying the authenticity of the LS. A notable number of
these studies were able to show that in quantum gravity (Planck scale) regime, there
is the possibility of violation of LS, see for instance \cite{C0}-\cite{S2}. This issue
reflects the fact that LS is not necessarily an exact symmetry of the nature in all energy
scales. Rather, it seems to be only an approximate symmetry governed on low energy scales, so that
in the Planck energy scale it loses credibility due to existence of a \emph{preferred reference
frame}. Indeed, the existence of a preferred state of motion leads
to a gross violation of the principle of relativity (that the laws of physics
are the same for all inertial observers) at the Planck energy scale. In other words, in Planck
scale there may be a reference frame in which the laws of physics might
appear to be different from those in other frames. From a cosmological viewpoint, it has an explicit
consequence that there may be a preferred cosmological rest frame
\cite{Sab}. However, along with violation of Lorentz invariance around Planck scale,
further conceptual challenges such as the invalidity of the equivalence principle at that energy scale arise automatically.
A common feature of different theories trying to provide a coherent description of Planck scale physics is that
the geometric structure of spacetime at this scale is noncommutative (NC) \cite{Dop1, Dop2}. In other words,
in Planck scale the NC field theory is governed as a framework of spacetime discreteness \cite{Douglas}.
In fact NC geometry is a powerful mathematical framework to describe a natural quantization of manifolds \cite{Ana}. Historically,
this issue for the first time was sparked by the Gelfand-Naimark theorem \cite{GN}.
Technically, this theorem expresses the fact that there is a one to one correspondence between
specific commutative algebras and specific spaces i.e. a duality between commutative
$C^{*}$ algebras and locally compact spaces. Indeed, this theorem indicates that
geometric structure of certain spacetimes can be seen within certain algebraic representations.
So, extension of commutative algebras to NC ones provides the NC algebraic structure on NC
spacetimes. Since the standard Minkowski spacetime at the quantum gravity
level becomes quantized, it seems to be natural that the commutative algebra of coordinates
$x^{\mu}$ on four dimensional real vector space (i.e. $[x^{\mu} , x^{\nu}]=0$)
are replaced by NC algebra $[x^{\mu}, x^{\nu}]\neq0$. Originally the NC algebraic relations
were formally introduced as $[x^{\mu} , x^{\nu}]=i\theta^{\mu\nu}$ where $\theta^{\mu\nu}$
denotes a constant $C$-number. These algebraic relations found considerable
popularity in quantum field theory (QFT) and also string theory. For instance, the seminal work of Seiberg
and Witten \cite{Witt} can be mentioned. They showed that in certain low energy limit
of open strings traveling in the background of a two form gauge field, the NC manifold
arises naturally. The important thing to note is that QFTs defined on the present form of NC
spacetime, do not meet the LS.  Latter on, the following Lie-algebraic
form of NC algebra with structure constants $\theta^{\mu\nu}_{\lambda}$ have been introduced
\cite{1}-\cite{9}
\begin{equation}
[x^{\mu}, x^{\nu}]=i\theta^{\mu\nu}_{\lambda}x^{\lambda}~.
\end{equation}
In fact, $\kappa$-Minkowski space or algebra is a restricted class of this algebra which obeys
the following commutation relations
\begin{equation}\label{e1-0}
[x^{0}, x^{i}]=\frac{i}{\kappa}x^{i},~~~[x^{0}, x^{0}]=[x^{i}, x^{j}]
=0,~~~~i,j=1,2,3
\end{equation}
Here, $x^{0}$ and $x^{i}$ signify the time and space operators, respectively.
For a detailed study on the mathematical formalism of $\kappa$-Minkowski
spacetimes along with some of its applications in Planck scale physics, see
\cite{Ana2}-\cite{12}. Many endeavors have been done to make a QFT relied strongly on a
$\kappa$-Minkowski spacetime. For instance one can mention \cite{QFT1}-\cite{QFT10}.
Unlike prior studies, in the present context one is not dealing with violation of LS.
On the other hand, Amelino-Camelia \cite{G} along with Magueijo and Smolin (MS) \cite{Smo}, independently proposed
alternative scenarios to special relativity (SR) which is called \emph{``doubly
special relativity"} (DSR). In these scenarios there is an extra invariant, \emph{Planck
length or energy} apart from \emph{the speed of light}. Note that both of these
scenarios address a type of QG with no imperil of the LS due to respecting
\emph{relativity of inertial frames}. Technically, in DSRs there is not necessarily
a breaking of LS, rather it contains just \emph{deformation} of this symmetry.
Interestingly, in Ref. \cite{J1} it has been shown that $\kappa$-Minkowski space is a realization
of the DSR, to the extent that thought to be one of the most prosperous possibilities
of DSR proposal, at least so far. More precisely, using the co-product of $\kappa$-Poincar\'{e}
algebra\footnote{Note that as authors of Ref. \cite{J1} have shown, there are infinitely many DSR constructions of the energy-momentum
and each of these constructions can be promoted to the $\kappa$-Poincar\'{e} quantum algebra.
along with the construction of $\kappa$-deformed phase space,
the NC spacetime structure can be extracted as well as the entire DSR phase space.}
In other words, the spacetime structure of the DSR proposal is equivalent to NC spacetime
one so that it suggests a NC version of Minkowski spacetime satisfying the LS.
Therefore, DSR as an effective approach to QG, expressly
predicts an energy cutoff $\kappa$ equivalent to an observer independent threshold length
scale. It is striking that from existence of such a threshold length scale
one can justify the lack of spontaneous formation of black holes in a
very condense district of spacetime \cite{Ghosh}. \\

By approaching the length scales comparable to $l_{p}$, since continuous metric idea fails,
one expects quantities such as the relativistic mass-shell condition (or dispersion
relation) generally gets modified to have the form such as
\begin{equation}\label{e1-1}
E^{2}=p^{2}+m^{2}+l_{p}E^{3}+...~.
\end{equation}
This general form of the modified dispersion relation
was suggested initially by Amelino-Camelia and Piran \cite{Pai}. Scenarios
with modified dispersion relations can be used in diverse areas of physics; to learn more about
some of these applications see for instance \cite{B1}-\cite{d3}. It is important to note that
in the context of DSR, addition of some \emph{nonlinear} terms to the Lorentz
transformations foils appearance of paradox between the existence of an observer independent threshold
length scale from one side and lack of Lorentz invariance of length in standard SR theory from other side,
see \cite{G, Smo} and also \cite{L1, L2, L3, L4} for a technical review. As a consequence of
nonlinear extension of the ordinary Lorentz group, one can say that in DSR framework spacetime
background coordinates are NC in essence. Therefore, to analyze the spacetime symmetries one should refer
to quantum groups (see \cite{re1}-\cite{re4} for a review of various aspects of DSRs in this respect). In the present
work, we focus mainly on the Magueijo-Smolin DSR proposal with assumption that the geometry of
Minkowski spacetime background deforms through the introduction of a NC geometry parameter $\kappa$.
As a result, the relativistic mass-shell condition can be modified as follows
\begin{equation}\label{e1-a}
E^{2}-p^{2}c^{2}=m^{2}c^{4}\Big(1-\frac{E}{\kappa}\Big)^{2}\;,
\end{equation}where $E$ and $p$ represent the magnitude of the energy and the three-momentum of the
particle with rest mass $m$, respectively\footnote{To give a quantitative estimation of the energy scale
needed to observe the expected experimental deviations from relativistic mass-shell relation, one may be
faced with the reasonable phenomenological question that whether the experimental predictions always
are confined to the unapproachable Planckian regime. Naturally deformed mass-shell condition
and subsequent experimental deviations should be valid and traceable at the "$\kappa$-scale" i.e.
QG scale. But, really at what scale QG effects become important? A general answer to this question is that
QG effects must be governed at scale $l_{QG}=1/M_{QG}$ (here $\hbar=1=c)$ so that $M_{QG}$ is expected
to be on the order of the Planck mass, $M_{Planck}$. It is more than a decade that serious
attempt have began to bounding $M_{QG}$ by measuring quantum gravitational effects through dispersion relations of
high energetic photons released from astrophysical sources such as gamma
ray bursts that can be detected by apparatus such as the Fermi telescope \cite{Fermi}. An explicit
bound reported in these studies is $M_{QG}>0.1M_{Planck}$; for details see the discussion raised by Amelino-Camelia and Smolin in \cite{2009}.}. It has been shown
that to preserve the form of the modified mass-shell condition in all inertial frames, the usual law of energy-momentum conservation
should be corrected too \cite{G}. Regarding the $\kappa$-Minkowski spacetime as made with NC coordinates
$x^{\alpha}$ that fulfill the Lie-algebra of the type (\ref{e1-0}), then Jacobi identity does not allow the
canonical commutation relations $\{x_{\mu},p_{\nu}\}=-g_{\mu\nu}$ to be unchanged. Therefore,
Poisson brackets between the coordinates of the phase space in the $\kappa$-Minkowski spacetime
obey the following modified algebra \cite{al1}-\cite{al4}
\begin{eqnarray}\label{e1-2}
\begin{array}{ll}
\{x_{\mu},x_{\nu}\}=\frac{1}{\kappa}(x_{\mu}\theta_{\nu}-x_{\nu}\theta_{\mu})~,\\\\
\{x_{\mu},p_{\nu}\}=-g_{\mu\nu}+\frac{1}{\kappa}\eta_{\mu}p_{\nu}~,\\\\
\{p_{\mu},p_{\nu}\}=0\;,
\end{array}
\end{eqnarray}
where $\mu,\nu=0,1,2,3$ and $\theta_{0}=1$, $\theta_{1,2,3}=0$. It is straightforward to check that
in the limit of $\kappa\rightarrow\infty$, this algebra reduces to the usual Minkowski Lie algebra
with the conventional canonical commutator relations between coordinates of the phase space. So far, we have
learned that the laws of DSR are nonlinear extension of the standard SR theory so that the variables
of the DSR are connected to its SR counterparts through a nonlinear mapping \cite{De1, De2}. In this respect,
some works have been developed in order to study mapping between $\kappa$-Minkowski
spacetime and its Minkowski counterpart through realization formalism, see for instance \cite{De3}-\cite{De7}.
We emphasize that for NC spaces there exist generally infinitely many
realizations in terms of commutative coordinates. Nevertheless, the physical outputs should be independent of these
realizations \cite{De7}. Note also that due to nontrivial change of the Poisson brackets (\ref{e1-2}), the mentioned map will be non-canonical.
In fact, the nonlinear mapping between DSR and SR never results in one to one correspondence between these two scenarios.
Rather, from a phenomenological perspective, DSR can provide a new extended framework independent of
SR one. In this respect we would stress that treatments which describe DSRs in terms of the maps from SR are usually misleading.
For instance some authors (e.g. \cite{sh}) based on such a treatment have claimed that DSRs \cite{G, Smo} are not
a new relativity.  Amelino-Camelia in \cite{Am} explicitly has responded to such a misconception. \\
Without getting into technical details, in what follows we just introduce nonlinear DSR Lorentz transformations
for our future purposes in this work (one can refer to \cite{De1, De2} for a detailed discussion). If we restrict the boost to the $x^{1}$ direction with
velocity $u^{1,2,3}=(u,0,0)$, then for 4-vectors $x^{\mu}=(t,\,\frac{x^{1,2,3}}{c})$ and $p^{\mu}=(\frac{E}
{c},\,p^{1,2,3})$, modified Lorentz transformations arising from NC geometry parameter $\kappa$ (as a realization of
Magueijo-Smolin proposal of DSR) are written as
\begin{equation}\label{e1-b}
t'=\gamma \alpha (t-\frac{u}{c^{2}}x^{1})~,~~~{x'}^{1}=\gamma \alpha (x^{1}-
\frac{u}{c^{2}}x^{0})~,~~~{x'}^{2}=\alpha x^{2}~,~~~{x'}^{3}=\alpha x^{3}~.
\end{equation}
and
\begin{equation}\label{e1-c}
{E'}=\frac{\gamma}{\alpha} (E-\frac{u}{c^{2}}p^{1})~,~~~{p'}^{1}=\frac{\gamma}
{\alpha}(p^{1}-\frac{u}{c^{2}}p^{0})~,~~~{p'}^{2}=\frac{p^{2}}{\alpha}~,~~~{p'}
^{3}=\frac{p^{3}}{\alpha}~.
\end{equation}
respectively, where $\gamma=(1-\frac{u^{2}}{c^{2}})^{-1/2}$ is the dimensionless
Lorentz contraction factor and $$\alpha=1+\frac{[(\gamma-1)p^{0}-\gamma up^{1}]}
{\kappa}.$$ It is obvious that in the limit of $\kappa\rightarrow\infty$ (i.e. in the
absence of an upper bound for energy scale), $\alpha\rightarrow1$ and the above
transformations reduce to the standard relativistic transformations. These
transformations show that the effects of NC geometry parameter is traceable via
nonlinear transformation rules \cite{sig}.

Despite the fact that DSRs suffer from the lack of a consistent
and well-established mathematical structure similar to what exists for SR,
$\kappa$-Minkowski spacetime and $\kappa$-Poincar\'{e} are likely the
richest frameworks which can be assigned to DSRs, at least so far.
Therefore, in the present paper, we just adopt the DSR interpretation proposed in
Magueijo-Smolin model \cite{Smo} as one of the conceivable phenomenological
frameworks of $\kappa$-Minkowski spacetime. As an interesting and promising
feature of the mentioned framework, it can be used to explain astrophysical data
received from GRBs (note that Gamma-ray bursts (GBRs) along with supernovae, neutron stars
and black holes are four main pillars of relativistic astrophysics) \cite{GRB}.

With these preliminaries and in the context of some prevalent phenomenological
issues through combination of the modified mass shell condition (\ref{e1-a})
and $\kappa$-deformed Lorentz transformations (\ref{e1-b}) and (\ref{e1-c}), we
examine and derive some phenomenological consequences of a $\kappa$-Minkowski
relativistic model. Specifically, we treat the elastic scattering problem
with the Planck energy cutoff by focusing on some relativistic kinematical
properties of this scattering problem. We are looking for the
mentioned objectives in these steps: In section 2 we study the effects of the
Planck energy upper bound, $\kappa$, on the elastic scattering process where
no new particles are created during the process. Section 3 is devoted to the
investigation of the \emph{rapidity} as one of the most important kinematical
variables involving in the heavy ion collisions in the presence of the $\kappa$
deformation of the Minkowski spacetime. The paper follows with a
conclusion along with a brief remark in section 4.


\section{$\kappa$-Deformed Elastic Scattering }

In this section, we investigate the elastic scattering processes where no new particles
are created during the process in a $\kappa$-deformed Minkowski space-time.
To this end, we firstly present an explicit relation for $\kappa$-deformed energy and
momentum. Suppose a clock that is fixed at the position $x_{1}$ of a rest frame, $S$. If the
clock emits signals in a regular time interval $\Delta t=t_{2}-t_{1}$, then according to the
modified Lorentz transformations (\ref{e1-b}), an observer located in the moving system $S'$,
measures this time interval as follows
\begin{equation}\label{e2-1}
\Delta t' =\gamma\alpha\left[\left(t_{2}-\frac{u}{c^{2}}x_{2}\right)-
\left(t_{1}-\frac{u}{c^{2}}x_{1}\right)\right]\;.
\end{equation}
Since the clock is fixed at the system $S$, then $x_{1}=x_{2}$, which
gives
\begin{equation}\label{e2-2}
\Delta t'=\alpha\gamma\Delta t\;.
\end{equation}
This equation can be rewritten in terms of the proper time $\tau$ as
\begin{equation}\label{e2-3}
\Delta t'=\alpha\gamma\Delta\tau\;.
\end{equation}
In this framework we have
\begin{equation}\label{e2-3}
p=m\frac{d\textbf{x}}{d\tau}=m\frac{d\textbf{x}}{dt}.\frac{dt}{d\tau}=\alpha\gamma mu\;,
\end{equation}
where $\textbf{u}=\frac{d\textbf{x}}{dt}$ is the classical velocity and $m$ marks the rest mass.
Now, inserting the $\kappa$-deformed relativistic momentum (\ref{e2-3}) into the modified dispersion relation
(\ref{e1-a}), we derive the following equation
\begin{equation}\label{e2-4}
\left(1-\frac{m^{2}c^{4}}{\kappa^{2}}\right)E^{2}+\frac{2m^{2}c^{4}}
{\kappa}E-m^{2}c^{4}\Big[1+\alpha^{2}(\gamma^{2}-1)\Big]=0\;.
\end{equation}
Solving this equation we find
\begin{equation}\label{e2-5}
E_{\kappa}= \frac{-\frac{2m^{2}c^{4}}{\kappa}\pm\left[\frac{4m^{4}c^{8}\alpha^{2}
(\gamma^{2}-1)}{\kappa^{2}}+4m^{2}c^{4}[1+\alpha^{2}(\gamma^{2}-1)]\right]^{1/2}}{2
\left(1-\frac{m^{2}c^{4}}{\kappa^{2}}\right)}\;.
\end{equation}
The positive sign in this solution is acceptable since for the case with $\kappa \rightarrow \infty$ and
$\alpha\rightarrow 1$, naturally the $\kappa$-deformed energy (\ref{e2-5}) should reduce to the relativistic
result $E=\gamma mc^{2}$. By applying the approximation $\frac{m^{2}c^{4}}{\kappa^{2}}\ll1$, Eq. (\ref{e2-5}) can be
rewritten as
\begin{equation}\label{e2-6}
E_{\kappa}= \frac{m^{2}c^{4}}{\kappa}\xi\;,
\end{equation}
where
\begin{equation}\label{e2-7}
\xi=-1+\sqrt{\alpha^{2}(\gamma^{2}-1)+\left(\frac{\kappa}{mc^{2}}\right)^{2}
[1+\alpha^{2}(\gamma^{2}-1)]}\;,
\end{equation}
is a dimensionless running constant. Note that $\lim_{\kappa\rightarrow\infty} E_{\kappa}=\gamma mc^{2}$.
In order to study the classical collision process, the center of mass (CM) coordinate system is an appropriate tools to
derive many of kinematic relations, while in the relativistic framework it is meaningless to speak about the CM system.
In SR theory, mass and energy are related so that it is common in SR kinematics that one uses a center of momentum
coordinate system instead of the CM. Of course, in the center of momentum coordinate as CM, the total linear momentum of the system is zero.
In the context of the relativistic collision, the laboratory coordinate system is related to the inertial
system $S$ and the center of momentum system $S'$ (the moving system) via a Lorentz transformation. In DSR framework we follow the
standard procedure with the difference that laboratory system $S$ and the moving system $S'$ are related now by the $\kappa$-deformed
Lorentz transformations. So, if a particle of rest mass $m_{1}$ which moves in one dimension collides elastically with
a particle of the rest mass $m_{2}$, then in the center of momentum system, we have\footnote{Here $p'_{1,2}$ denote
the magnitude of the 3-momentum of particles $m_{1,2}$ in the CM system so that $\textbf{p}'_1+\textbf{p}'_2=0$.
An observer located in the CM frame sees these two particles are moving towards each other so that depending on the direction, one has $p'_1-p'_2=0$ or $p'_2-p'_1=0$.}
\begin{equation}\label{e2-8}
p'_{1}= p'_{2}\;.
\end{equation}
At the first glance this equality seems to be misleading. It is obvious that
from the deformed Lorentz transformations one naturally gets the deformed dispersion relation (\ref{e1-a}) (since this transformation
keeps (\ref{e1-a}) invariant). Therefore, one has also the deformed conservation of energy-momentum so that
this relation then governs on the nature of how particles behave in collisions. More precisely,
one expects that in the standard momentum conservation law (\ref{e2-8}), $\kappa$-deformation must be regarded.
While this is sensible in essence, as we show equation (\ref{e2-8}) is still valid up to the first order correction.
\\
By using the relation $\beta\gamma=\sqrt{\gamma^{2}-1}$, the space components of the
momentum 4-vector in (\ref{e2-8}) can be written as
\begin{equation}\label{e2-8*}
m_{1}c\alpha \sqrt{\gamma'^{2}_{1}-1}= m_{2}c\alpha \sqrt{\gamma'^{2}_{2}-1}\;,
\end{equation}
since $p_1=\alpha\gamma_{1} m_{1}u_{1} $,~ $p_2=\alpha\gamma_{2} m_{2}u_{2} $
and also $\beta\equiv \frac{u}{c}$. According to the deformed Lorentz transformation
(\ref{e1-c}) and using (\ref{e2-6}), the transformation of the momentum $p_1$ (from $S$ to $S'$)
is written as
\begin{equation}\label{e2-9*}
p'_1= m_{1}c\beta_{1}\gamma'_{1}\gamma'_{2}-\frac{m_{1}^{2}c^{3}}{\kappa\alpha}\beta'_{2}\gamma'_{2}\xi_{1}\;.
\end{equation}
Finally by replacing this relation into the left hand side of relation (\ref{e2-8*}), we arrive at the following relation
\begin{equation}\label{e2-9}
m_{1}c\beta_{1}\gamma'_{1}\gamma'_{2}-\frac{m_{1}^{2}c^{3}}{\kappa\alpha}\beta'_{2}\gamma'_{2}\xi_{1}=
m_{2}c\alpha\sqrt{\gamma'^{2}_{2}-1}\;,
\end{equation}
where this relation can be solved for $\gamma'_{1}$ and $\gamma'_{2}$ in terms of $\gamma_{1}$ to find
\begin{equation}\label{e2-10}
\gamma'_{1}=\frac{\left(\frac{m_{1}}{m_{2}}\alpha+\frac{m_{1}c^{2}}
{\kappa\alpha}\xi_{1}\right)}{\sqrt{(1-\gamma^{2}_{1})+\left(\frac{m_{1}}{m_{2}}
\alpha+\frac{m_{1}c^{2}}{\kappa\alpha}\xi_{1}\right)^{2}}}\;,
\end{equation}
and
\begin{equation}\label{e2-11}
\gamma'_{2}=\frac{\left(\frac{m_{2}}{m_{1}}\alpha+\frac{m_{1}c^{2}}
{\kappa\alpha}\xi_{1}\right)}{\sqrt{(1-\gamma^{2}_{1})+\left(\frac{m_{2}}{m_{1}}
\alpha+\frac{m_{1}c^{2}}{\kappa\alpha}\xi_{1}\right)^{2}}}\,.
\end{equation}
In the absence of the $\kappa$ deformation of the Minkowski space-time background i.e.
for $\kappa\rightarrow\infty$, then $\alpha=1$ and  $\frac{m_{01}c^{2}}{\kappa\alpha}\xi_{1}=
\gamma_{1}$. So, the above equations reduce to the flowing special relativistic counterparts
\begin{equation}\label{e2-12}
\gamma'_{1}=\frac{\left(\frac{m_{1}}{m_{2}}+\gamma_{1}\right)}
{\sqrt{1+2\gamma_{1}(\frac{m_{1}}{m_{2}})+(\frac{m_{1}}{m_{2}})^{2}}}\;,
\end{equation}
and
\begin{equation}\label{e2-13}
\gamma'_{2}=\frac{\left(\frac{m_{2}}{m_{1}}+\gamma_{1}\right)}
{\sqrt{1+2\gamma_{1}(\frac{m_{2}}{m_{1}})+(\frac{m_{2}}{m_{1}})^{2}}}\;.
\end{equation}
Now we write the transformation equations for momentum
components between the moving system $S'$ and the laboratory system $S$ after the scattering.
It is obvious that after scattering we have both $x$ and $y$ components of the momentum.
The $x$ and $y$ components of the momentum after scattering in the laboratory system $S$
read as follows
\begin{equation}\label{e2-14}
p_{1,x}=m_{1}c\gamma'_{2}\left(\beta_{1}\alpha\gamma'_{1}\cos\theta+
\frac{m_{1}c^{2}}{\kappa\alpha}\beta'_{2}\xi'_{1}\right)\;,
\end{equation}
and
\begin{equation}\label{e2-15}
p_{1,y}=m_{1}c\beta'_{1}\gamma'_{1}\alpha\sin\theta\;.
\end{equation}
By introducing the angle of scattering $\psi$ in the laboratory frame and by dividing
(\ref{e2-15}) with (\ref{e2-14}), we get
\begin{equation}\label{e2-16}
\tan\psi=\frac{\alpha\sin\theta}{\gamma'_{2}\left(\alpha\cos\theta+
\frac{m_{1}c^{2}}{\kappa\alpha\gamma'_{1}}(\frac{\beta'_{2}}{\beta'_{1}})\xi'_{1}\right)}\;,
\end{equation}
where by inserting $p'_{1}=p'_{2}$, it takes the following form
\begin{equation}\label{e2-17}
\tan\psi=\frac{\alpha\sin\theta}{\gamma'_{2}\left(\alpha\cos\theta+
\frac{m_{1}c^{2}}{\kappa\alpha\gamma'_{1}}(\frac{m_{1}\gamma'
_{1}}{m_{2}\gamma'_{2}})\xi'_{1}\right)}\;.
\end{equation}
By the same procedure, for the recoiled particle we find respectively
\begin{equation}\label{e2-18}
p_{2,x}=m_{2}c\gamma'^{2}_{2}\beta'\left(\frac{m_{2}c^{2}}
{\kappa\alpha\gamma'_{2}}\xi'_{2}-\cos\theta\right)\;,
\end{equation}
and
\begin{equation}\label{e2-19}
p_{2,y}=-m_{2}c\beta'_{2}\gamma'_{2}\alpha\sin\theta\;.
\end{equation}
Introducing an angle of scattering $\eta$ in the laboratory frame for the recoiled
particle and then by dividing (\ref{e2-19}) with (\ref{e2-18}), we find
\begin{equation}\label{e2-20}
\tan\eta=-\frac{\alpha\sin\theta}{\gamma'_{2}\left( \frac{m_{2}
c^{2}}{\kappa\alpha\gamma'_{2}}\xi'_{2}-\cos\theta\right)}\;.
\end{equation}
For the special case with $ m_{1}=m_{2}$, i.e. with
\begin{equation}\label{e2-20m}
\gamma'_{1}=\gamma'_{2}=\frac{1}{f(\gamma_{1})}\;,
\end{equation}
we find
\begin{equation}\label{e2-21}
f(\gamma_{1})=\sqrt{1+\frac{\alpha^{2}(1-\gamma^{2}_{1})}{\left(\alpha^{2}+\sqrt{1+
\alpha^{2}(\gamma^{2}_{1}-1)}\right)^{2}}}\;.
\end{equation}
Therefore, for the case $ m_{1}=m_{2}$, the angles $\psi$ and $\eta$ take the following
forms respectively
\begin{equation}\label{e2-22}
\psi=\arctan\left(\frac{\alpha\sin\theta}{\frac{\alpha\cos\theta}
{f(\gamma_{1})}+\sqrt{\frac{1}{f^{2}(\gamma_{1})}+\frac{1}{\alpha^{2}}-1}}\right)\;,
\end{equation}
and
\begin{equation}\label{e2-23}
\eta=\arctan\left(\frac{\alpha\sin\theta}{-\frac{\alpha\cos\theta}
{f(\gamma_{1})}+\sqrt{\frac{1}{f^{2}(\gamma_{1})}+\frac{1}{\alpha^{2}}-1}}\right)\;,
\end{equation}
So, the angle between the directions of the scattered and recoiled particles is given by
$\phi=\psi+\eta$. Through the relation $\frac{d\phi}{d\theta}=0$, one finds that for $\theta
=\pm\frac{\pi}{2}$ the angle $\phi$ has the maximum
\begin{equation}\label{e2-24}
\phi_{\kappa,Max}=2\arctan\left(\frac{\alpha}{\sqrt{\frac{1}{f^{2}(\gamma_{1})}+\frac{1}
{\alpha^{2}}-1}}\right)
\end{equation}
As expected, for the case $\alpha\rightarrow1$, Eq. (\ref{e2-24}) recovers its relativistic
counterpart as
\begin{equation}\label{e2-25}
\phi_{SR,Max}=2\arctan\sqrt{\frac{2}{1+\gamma_{1}}}.
\end{equation}
Therefore, one finds that in the $\kappa$-deformed non-commutative geometry, the maximum
amount of the included angle $\phi$ is dependent on the two dimensionless parameters $\gamma_{1}$
and $\alpha$. In order to have a qualitative understanding of Eq. (\ref{e2-24}), we plot
variation of $\phi_{Max}$ in terms of $\gamma_{1}$ in figure 1. If we set the $\alpha$ to be a
running constant, this figure gives a qualitative
description of the situation. For any given value of $\gamma_{1}$, by increasing the value of $\alpha$ the
maximum included scattering angle $\phi_{Max}$ increases. So, we can say that correction of $\alpha$ due to
the $\kappa$-deformed Minkowski spacetime results in a shift of the maximum included scattering
angle $\phi_{Max}$.
\begin{figure}[htbp]
 \centering\includegraphics[width=3.5in]{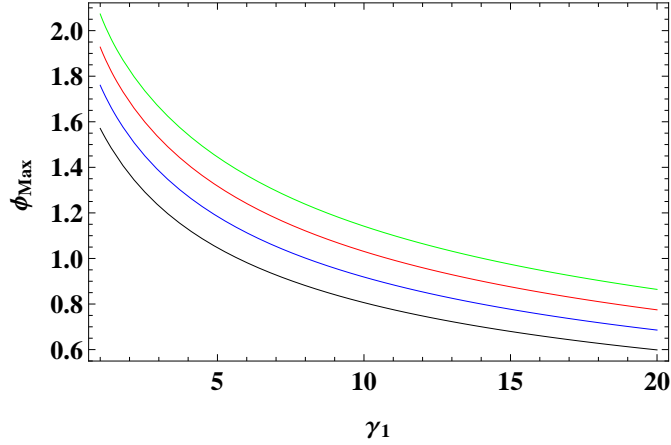}
\caption{ \small {Variation of the maximum included scattering angle $\phi_{Max}$ in terms of the
dimensionless Lorentz contraction factor $\gamma_{1}$, for different values of $\alpha$ due to
$\kappa$-deformed Lorentz transformation: $\alpha=1$ (Black), $\alpha=1.1$ (Blue),
$\alpha=1.2$ (Red) and $\alpha=1.3$ (Green). The unit of $\phi$ is in radian.}}
 \label{dbcontour}
 \end{figure}

In what follows, to see another prediction of the setup, we assume that the recoil angle $\eta$
is negligible. Then we track the effect of $\kappa$ modification on the kinetic energy $K$
of the scattered particle 1  by the target particle 2. Using the inverse $\kappa$-modified Lorentz transformation
(\ref{e1-c}), we have
\begin{equation}\label{e2-26}
E_{1,a}=\frac{\gamma'_{1}}{\alpha}\Big(E'_{1,a}+c \beta'_{1} p'_{1,x}
\cos\theta\Big)\;,
\end{equation}
where $E_{1,a}$ and $E'_{1,a}$ represent the total energy of particle 1 after collision
in the laboratory and center of momentum frames, respectively. By substituting $E'_{1,a}=mc^{2}
\gamma'_{1}$ and $p'_{1,a}=mc \beta'_{1}\gamma'_{1}$ into
Eq. (\ref{e2-26}), we get
\begin{equation}\label{e2-27}
E_{1,a}=mc^{2}\frac{\gamma'^{2}_{1}}{\alpha}(1+\beta'^{2}_{1}
\cos\theta)\;.
\end{equation}
By introducing $K_{1,b}$ and $K_{1,a}$ as the kinetic energies of the particle 1
before and after scattering respectively, then one can write
\begin{equation}\label{e2-28}
\frac{K_{1,a}}{K_{1,b}}=\frac{\frac{\gamma'^{2}_{1}}{\alpha}
+\frac{\gamma'^{2}_{1}-1}{\alpha}\cos\theta-1}{\gamma_{1}-1}\;,
\end{equation}
Using equation (\ref{e2-20m}), this equation can be rewritten as follows
\begin{equation}\label{e2-29}
\frac{K_{1,a}}{K_{1,b}}=\frac{\frac{1}{\alpha f^{2}(\gamma_{1})}+
\frac{1}{\alpha}(\frac{1}{f^{2}(\gamma_{1})}-1)\cos\theta-1}
{\gamma_{1}-1}\;.
\end{equation}
Now we introduce a scattering angle $\theta$ in the center of
momentum frame in terms of the $\psi$ in the laboratory frame. For this purpose, by
squaring Eq. (\ref{e2-22}), we find
\begin{equation}\label{e2-30}
A_{1}\cos^{2}\theta+A_{2}\cos\theta+A_{3}=0\;,
\end{equation}
where by definition
\begin{eqnarray}\label{e2-31}
\begin{array}{ll}
A_{1}\equiv\alpha^{2}\big(1+\frac{\tan^{2}\psi}{f^{2}(\gamma_{1})}\big)~,\\\\
A_{2}\equiv2\tan^{2}\psi\sqrt{\frac{\alpha^{2}}{f^{4}(\gamma_{1})}+\frac{(1-\alpha^{2})}{f^{2}(\gamma_{1})}}~,\\\\
A_{3}\equiv\tan^{2}\psi\big(\frac{1}{f^{2}(\gamma_{1})}+\frac{1}{\alpha^{2}}-1\big)-\alpha
^{2}\;.
\end{array}
\end{eqnarray}
The solution of equation (\ref{e2-30}), after some rearrangement, reads as follows
\begin{eqnarray}\label{e2-32}
\cos\theta&=&\frac{-\frac{\tan^{2}\psi}{f^{2}(\gamma_{1})}\sqrt{\frac{1}{\alpha^{2}}+\frac{f^{2}
(\gamma_{1})(1-\alpha^{2})}{\alpha^{4}}}}{1+\frac{\tan^{2}\psi}{f^{2}(\gamma_{1})}}\nonumber\\ & &
\pm \frac{1}{4[\alpha^{4}\sin^{2}\psi+\alpha^{2}f^{2}(\gamma_{1})\cos^{2}\psi]}\times \nonumber\\ & &
\left\{\Big[4\alpha^{6}f^{2}(\gamma_{1})-8\alpha^{8}f^{2}(\gamma_{1})+4\alpha^{4}f^{4}(\gamma_{1})
+4\alpha^{8}f^{4}(\gamma_{1})\Big]\right. \nonumber\\ & &\left.
\cos^{2}(2\psi)+  8\alpha^{8}f^{4}(\gamma_{1})\cos(2\psi)-4\alpha^{6}f^{2}(\gamma_{1}) \right. \nonumber\\ & &\left.
+4\alpha^{8}f^{2}(\gamma_{1})-4\alpha^{4}
f^{4}(\gamma_{1})+4\alpha^{6}f^{4}(\gamma_{1})+\right. \nonumber\\ & &\left.
4\alpha^{8}f^{4}(\gamma_{1})\right\}^{1/2}
\end{eqnarray}
Through a straightforward calculation, one can show that equation (\ref{e2-32}) in the limit of $\alpha\rightarrow 1$
recovers its special relativistic counterpart as follows
 \begin{equation}\label{e2-33}
\cos\theta=\frac{-\frac{(\gamma_{1}+1)}{2}\tan^{2}\psi\pm1}{1+
\frac{(\gamma_{1}+1)}{2}\tan^{2}\psi}\,.
\end{equation}
Finally, by inserting this relation along with (\ref{e2-21}) into equation (\ref{e2-29}), we are able to show the
qualitative behavior of $\frac{K_{1,a}}{K_{1,b}}$ in terms of the laboratory scattering angle $\psi$, as shown in figure 2.
\begin{figure}[htbp]
 \centering\includegraphics[width=3.5in]{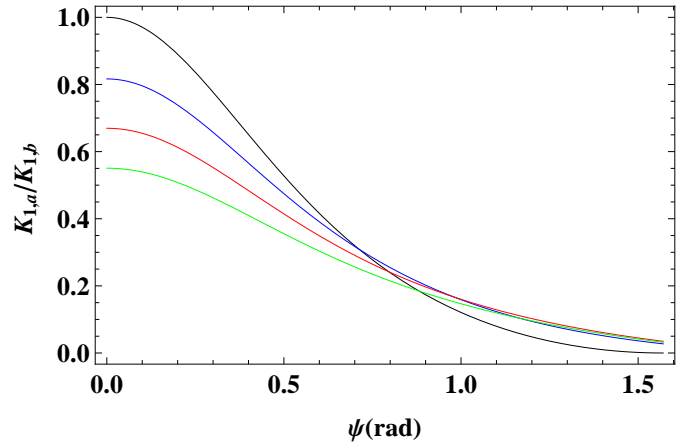}
\caption{ \small{ Variation of $\frac{K_{1,a}}{K_{1,b}}$ in terms of Lab scattering
angle $\psi$ for different values of $\alpha$ due to
$\kappa$-deformed Lorentz transformation: $\alpha=1$ (Black), $\alpha=1.1$
(Blue), $\alpha=1.2$ (Red) and $\alpha=1.3$ (Green).
We also set the dimensionless relativistic parameter $\gamma=10$ which is equivalent to
$u\approx 0.99 c$. The unit of $\psi$ is in radian. }}
 \label{dbcontour}
 \end{figure}
As we see in this figure, contrary to our expectation based on the SR theory
($\alpha=1$) and even classical mechanics\footnote{In the framework of Newtonian
kinematics and for the case of two equal masses $m_{1}=m_{2}$, the ratio $\frac{K_{1,a}}{K_{1,b}}$ in terms of
$\psi$ results in the simple relation $\frac{K_{1,a}}{K_{1,b}}=\cos^{2}\psi$ which the laboratory scattering angle $\psi$
is limited to the interval $0\leq\psi\leq\frac{\pi}{2}$. For more details one can
see ``Classical Mechanics" text book such as Ref. \cite{Text}. If $\psi=0$ then $\frac{K_{1,a}}{K_{1,b}}=1$, namely,
no collision takes place since the speed remains constant before and after the collision and no momentum is
transferred to particle-2. However, if $\psi\rightarrow\frac{\pi}{2}$ then $K_{1,a}\rightarrow0$
($K_{2,a}=K_{1,b}$) and particle-2 is forwardly scattered with the same kinetic energy before owned by
particle-1. }
(that is, $\alpha=1=\gamma$, in Eqs. (\ref{e2-29})
and (\ref{e2-33})), for the case $\psi=0$ we have $\frac{K_{1,a}}{K_{1,b}}<1$. Here,
we are confronted with an unusual status in which as soon as we turn on $\kappa$-spacetime
deformation, the nature of two-body collision alters from elastic to inelastic one. It
is important to stress that the overall behavior of the curves in figure 2 are
independent of the values of the dimensionless relativistic parameter $\gamma$.\\
At this point a question arises whether it is possible to remove these unusual
phenomenological effects via $\kappa$ deformation of the momentum conservation law (\ref{e2-8}).
To check this situation, inspired by \cite{Smo} we modify Eq. (\ref{e2-8}) as follows
\begin{equation}\label{e2-34}
\frac{p'_{1}}{1-\frac{\alpha}{\kappa}E'_{1}}=\frac{p'_{2}}{1-\frac{\alpha}
{\kappa}E'_{2}}~.
\end{equation}
Let's focus carefully on this modified
momentum conservation law in the CM system. Since in the adopted DSR proposal (the MS proposal)
one has the spacetime translational invariance, the energy and momentum are conserved.
However, DSR proposals are non-linear in essence. This means that
4-momentum of a system of two particles is not just a linear
summation of two particles momentums. As a consequence,
in passing from SR to DSR, i.e. from linear to non-linear relativity,
the additivity of energy and momentum is replaced with non-additivity character.
For more insight on this issue we refer the reader to \cite{Smo}.
Now the modified relation (\ref{e2-34}) can be rewritten as follows
\begin{equation}\label{e2-35}
p'_{1}=\bigg(1+\frac{\alpha m^{2}c^{4}}{\kappa^{2}}(\xi'_{2}-\xi'_{1})-\Big(\frac{\alpha
m^{2}c^{4}}{\kappa^{2}}\Big)^{2}\xi'_{1}\xi'_{2}\bigg)p'_{2}~,
\end{equation}
where $m_{1}=m_{2}=m$. Up to the first order of the modification
(i.e. $\kappa^{-1}$), the second and third terms on the
right hand side of this relation can be neglected. This means that, up to
the first order correction, momentum conservation law (\ref{e2-8}) is still valid
and consequently the obtained unusual outcomes are direct consequences
arising from extended phenomenological framework at hand.
\\
As the final remark in this section, we note that
the apparent energy non-conservation appeared above has been conjectured in NC
manifolds and could be due to a discrete (underlying) time evolution \cite{Bala}. So, Noether's
theorem should be correspondingly reformulated for the case of NC spacetime
translations.


\section{$\kappa$-deformed rapidity }

In this section we firstly present an important kinematical variable, the rapidity,
that relates the momentum of the particle to the dynamics of a
heavy-ion reaction. There is the possibility of creation of several particles after collision. The momentum
of each particle can be decomposed into a longitudinal component $p_{l}$ and a transverse
component $p_{t}$ with reference to the collision axis. The longitudinal momentum of a
particle, due to its dependence on the velocity of the CM frame with respect to the laboratory
frame, is not so conventional variable in literature. Besides, in order to analyze some
experimental outputs, it is essential to be able to view them from the CM frame's perspective.
As has been mentioned previously, we use the word "CM frame" for center of momentum frame in this context.
By introducing a kinematical variable as the rapidity $y$, one is able simply to choose or change the reference frame.
This is arising from the fact that unlike the velocity, the variable $y$ is defined in special relativity to be additive under a sequence of the Lorentz
transformations along the same direction. In what follows, we answer the question whether
the variable $y$ still remains additive under successive $\kappa$-deformed Lorentz transformations
(\ref{e1-c}). In three dimensional space, modified on shell condition (\ref{e1-a}), can be rewritten as follows
\begin{equation}\label{e3-1}
E^{2}=p^{2}_{l}+M^{2}_{t}\;,
\end{equation}
so that
\begin{equation}\label{e3-2}
M^{2}_{t}=p^{2}_{t}+m^{2}\Big(1-\frac{E}{\kappa}\Big)^{2}\;,
\end{equation}
is known as transverse mass. By writing Eq. (\ref{e3-1}) as $(\frac{E}{M_{t}})^{2}-(\frac{p_{l}}
{M_{t}})^{2}=1$, then the variable $y$ can be defined in terms of the energy and momentum
as follows
\begin{equation}\label{e3-3}
E=M_{t}\cosh y,\quad p_{l}=M_{t}\sinh y\;.
\end{equation}
Therefore, for a $\kappa$-deformed special relativity scenario
with dispersion relation as Eq. (\ref{e1-a}), one finds that the rapidity/energy-momentum relation
reads as Eq. (\ref{e3-3}) which is similar to its relativistic counterpart.
So, the relation between velocity and rapidity now is obtained from Eq. (\ref{e3-3}) as
\begin{equation}\label{e3-4}
u_{l}\equiv\frac{cp_{l}}{E}=c\tanh y\,.
\end{equation}
This relation gives the rapidity $y$  as follows
\begin{equation}\label{e3-5}
y=\frac{1}{2}\ln\left(\frac{1+u_{l}}{1-u_{l}}\right)=
\frac{1}{2}\ln\left(\frac{E+p_{l}}{M_{t}}\right)\;,
\end{equation}
since
\begin{equation}\label{e3-6}
\tanh^{-1}z=\frac{1}{2}\ln\left(\frac{1+z}{1-z}\right)\;.
\end{equation}
To investigate the additivity of rapidity under a sequence of the $\kappa$-deformed Lorentz transformations,
we consider the transformation of the momentum vector under a change
of the reference frame along the collision axis. It is obvious that under such a transformations, $p_{t}$
and $M_{t}$ are unchanged. Therefore, under a $\kappa$-deformed Lorentz transformation, the energy and
longitudinal component of the momentum transform as
\begin{equation}\label{e3-7}
E'=\frac{\gamma_{c}}{\alpha}\Big(E+\frac{u_{c}}{c^{2}}p_{l}\Big),\quad\quad {p'}_{l}=
\frac{\gamma_{c}}{\alpha}\Big(p_{l}+\frac{u_{c}}{c^{2}}E\Big)\;.
\end{equation}
Note that here a prime marks the quantities that are measured by an observer in the laboratory system
which moves with the velocity $u_{c}$ with respect to the CM frame of reference in which the energy $E$
and momentum $p_{l}$ are measured. Putting Eq. (\ref{e3-1}) along with $\gamma_{c}=\cosh y_{c}$ and $
\gamma_{c}u_{c}=\sinh y_{c}$ derived from Eq. (\ref{e3-4}) into Eq. (\ref{e3-7}), we get
\begin{equation}\label{e3-8}
E'=\frac{M_{t}}{\alpha}\cosh(y+y_{c}),\quad\quad  {p'}_{l}=\frac{M_{t}}{\alpha}
\sinh(y+y_{c})\;.
\end{equation}
Therefore, the rapidity $y'$ as seen in the laboratory frame can be read as follows
\begin{equation}\label{e3-9}
y'=\cosh^{-1} \left(\frac{\cosh(y+y_{c})}{\alpha}\right)\,.
\end{equation}
While this relation reflects the fact that rapidity is not an additive quantity under a sequence of the
$\kappa$-deformed Lorentz transformations, for the case $\alpha=1$ (that is, under standard Lorentz transformations)
this relation restores the additivity property, i.e, $y'=y+y_{c}$. Therefore, in passing from the standard SR to
$\kappa$-deformed SR, one loses the additivity nature of rapidity.
This is much similar to losing additivity rule of speeds in passing from Galilean relativity to special relativity.
Note that relation (\ref{e3-9}) is a direct result of the nonlinear representation of Lorentz transformations.
Technically, as is shown in Ref. \cite{De2}, the Magueijo-Smolin proposal of DSR can be
realized via deformed translation invariance without need to deformation of the LS (this is also the case
for all other proposals of DSR, see \cite{al4}). The $\kappa$-Minkowski spacetime modified translation generator
in the MS model of DSR is given by $t^{\mu}=\frac{p^{\mu}}{1-\alpha p/\kappa}$. In this respect, the origin of deviations from SR
observed in this paper comes back to the deformed translation invariance since this
really controls how momentum of particles behaves in collisions.


\section{Conclusion}

Noncommutative geometry (spacetime) is undoubtedly
one of the richest framework among other alternatives to pursue
physics at the Planck scale. In this paper, we were concerned on a
particular form of noncommutative spacetimes which highly takes care of
Lorentz Symmetry and known as $\kappa$-Minkowski
spacetime. The noncommutative parameter $\kappa$ can be conceived
as quantum gravity scale where high energy physics is trying
to unravel its mysteries. The mentioned noncommutative parameter is
observer independent which leads to a natural connection
between $\kappa$-Minkowski spacetime and Doubly Special
Relativity theories (DSRs). Therefore, noncommutative geometry parameter
$\kappa$ is responsible for cutoff energy scale featured in
DSR theories \cite{G, Smo} as the seconded invariant. In the new framework,
we are faced with DSR formalism as a nonlinear $\kappa$ extension
of the Lorentz transformations and dispersion relations of Special
Relativity. In this work, through focusing on the Magueijo-Smolin
version of DSR theories \cite{Smo}, we have investigated how such energy
upper bound will affect some relativistic kinematical parameters in a
typical collision of particles. As a first step, by applying $\kappa$-deformed
Lorentz transformations (\ref{e1-b}), (\ref{e1-c}) along with  modified
mass shell condition (\ref{e1-a}), we have investigated the effect of the
Planck energy cutoff on the elastic scattering processes for the case that
no new particles are produced in the process. We have shown (as figure 1
indicates) that the existence of $\alpha$-term caused by the $\kappa$-Minkowski
spacetime, leads increment of maximum included scattering
angle $\phi_{Max}$ relative to the standard case. While the classical and special relativistic kinematics
indicate that the maximum included scattering angle is bounded as $\phi_{Max}\leq\frac{\pi}{2}$, the
$\kappa$-deformed nonlinear extension of special relativity violates this bound since for
angles larger than $\frac{\pi}{2}$, there is probability
of detection of scattered particles. In the same way, by assuming that the
recoil angle $\eta$ is negligible, we have treated the effect of the Planck energy upper bound
($\kappa$ modification) on the kinetic energy $K$ of the scattered particle.
Surprisingly, in contrast to classical mechanics and even
special theory of relativity, we have observed that for the case $\psi=0$ and $\alpha>1$,
$\frac{K_{1,a}}{K_{1,b}}<1$ as figure 2 shows. Considering the fact that for
the case $\alpha=1$ one recovers the standard result $\frac{K_{1,a}}
{K_{1,b}}=1$, in the context of the $\kappa$ deformed extension of the
special relativity one tempts to think that spacetime seems to have some sort of dissipative effects at quantum gravity scale.
The apparent energy non-conservation appeared in our context could be due to a discrete (underlying) time
evolution as shown in Ref. \cite{Bala}. It is also expected that the Noether's theorem must be correspondingly
reformulated in the present case of NC spacetime translations.
As an important result, we have shown that unlike the special relativity, rapidity in $\kappa$-Minkowski spacetime is no longer an additive quantity under a sequence
of $\kappa$-deformed Lorentz transformations. This is reminiscent of the fact that one loses the additive
nature of speeds in passing from Galilean relativity to special relativity.
We note that the observed deviations from SR in this paper originate from
the deformed translation invariance, not necessarily to the deformed LS. In fact, this deformed
translation invariance is that controls how momentum of particles behaves in collisions. The outputs of our study support the idea that
the main feature of DSR theories is the deformed translation symmetry, not the deformed LS as has been discussed also in \cite{De2, al4}.

Finally, about the algebraic structure of $\kappa$-Minkowski spacetime if
we look at the full algebra generated by boosts, rotations
and translations, then the algebra is really the Lorentz algebra. Depending on the choice
of realization, even the translation part can be un-deformed and therefore having the full Poincar\'{e} algebra. But
this is only on the algebra level, that is what governs the one particle representations.
However the coalgebraic sector is deformed, leading to nontrivial multi-particle states, which are the ones appearing
in collisions. The $\kappa$-Minkowski algebra is compatible with deformed Poincare algebra,
usually $\kappa$-Poncar\'{e} algebra (but one can also go beyond this in general to deformations of igl-Hopf
algebra) for which we have that the Lorentz sector is intact, but depending on realization the translation
part is deformed. Only in the classical basis (or natural realization) the translation part is also intact.
Anyway, for any realization (natural, bicrossproduct etc)
the coalgebraic sector is deformed, meaning that the coproduct of boosts, rotation and translation is not
primitive, and this gives some new interesting phenomenon for the multiparticle states, and therefore collisions \cite{Tnx}.\\


{\bf Acknowledgement}\\
We are indebted to two anonymous referees for
detailed and very insightful comments.


\begin{thebibliography}{99}
\bibitem{C0}
S. M. Carroll, G. B. Field
and R. Jackiw, Phys. Rev. D
41, 1231 (1990)
\bibitem{C1}
S. Coleman and S. L. Glashow,
Phys. Lett. B \textbf{405}, 249
(1997)
\bibitem{C2}
D. Colladay and V. A. Kostelecky,
Phys. Rev. D \textbf{55}, 6760 (1997)
\bibitem{C3}
R. M. Mansouri, R. U. Sexl, Gen. Rel.
Grav. \textbf{8}, 497 (1977); ibid. \textbf{8}, 515
(1977)
\bibitem{C4}
J. A. Lipa, J. A. Nissen, S. Wang,
D. A. Stricker, D. Avaloff, Phys.
Rev. Lett. \textbf{90}, 060403 (2003).
\bibitem{A0}
G. Amelino-Camelia, J. Ellis, N. E.
Mavromatos and D. V. Nanopoulos, Int.
J. Mod. Phys. A \textbf{12}, 607 (1997)
\bibitem{A1}
G. Amelino-Camelia,
J. Ellis, N. E. Mavromatos, D. V.
Nanopoulos and S. Sarkar,  Nature \textbf{393}
763 (1998)
\bibitem{A2}
S. D. Biller et al, Phys. Rev. Lett.
\textbf{83}, 2108 (1999)
\bibitem{A3}
T. Kifune, Astrophys. J. Lett.
\textbf{518}, L21 (1999)
\bibitem{A4}
R. Aloisio, P. Blasi, P. L. Ghia and A.
F. Grillo, Phys. Rev. D \textbf{62}, 053010 (2000)
\bibitem{A5}
R. J. Protheroe and H. Meyer, Phys. Lett.
B \textbf{493}, 1 (2000)
\bibitem{A6}
J. Alfaro, H. A. Morales-Tecotl and L. F.
Urrutia, Phys. Rev. Lett. \textbf{84}, 2318 (2000)
\bibitem{A7}
G. Amelino-Camelia and
T. Piran, Phys. Rev. D \textbf{64},
036005 (2001)
\bibitem{A8}
G. Amelino-Camelia,
Nature \textbf{408}, 661 (2000).
\bibitem{S0}
S. Sarkar, Mod. Phys. Lett.
A \textbf{17}, 1025 (2002)
\bibitem{S1}
D. V. Ahluwalia, Mod. Phys. Lett.
A \textbf{17}, 1135 (2002)
\bibitem{S2}
T. Jacobson, S. Liberati and D.
Mattingly, Phys. Rev. D \textbf{66}, 081302
(2002)
\bibitem{Sab}
S. Hossenfelder, L. Smolin,
Phys. Canada \textbf{66}, 99 (2010)
\bibitem{Dop1}
S. Doplicher, K. Fredenhagen, J. E.
Roberts, Phys. Lett. B \textbf{331}, 39 (1994)
\bibitem{Dop2}
S. Doplicher, K. Fredenhagen, J. E. Roberts,
Commun. Math. Phys. \textbf{172}, 187 (1995)
\bibitem{Douglas}
M. R. Douglas, and N. A. Nekrasov,
Rev. Mod. Phys. \textbf{73}, 977 (2001).
\bibitem{Ana} A. Pacho{\l}, Journal of
Physics: Conference Series \textbf{442}, 012039
(2013)
\bibitem{GN}
I. M. Gelfand, M. A. Naimark, Math.
Sbornik \textbf{12}, 2 (1943)
\bibitem{Witt}
N. Seiberg and E. Witten, JHEP \textbf{9909}, 032
(1999)
\bibitem{1}
J. Lukierski, H. Ruegg, Phys. Lett. B \textbf{329}
, 189 (1994)
\bibitem{2}
J. Lukierski, H. Ruegg, A. Nowicki, V. N. Tolstoi
Phys. Lett. B \textbf{264}, 331 (1991)
\bibitem{3}
S. Majid, H. Ruegg, Phys. Lett. B \textbf{334},
348 (1994)
\bibitem{4}
J. Lukierski, A. Nowicki, H. Ruegg,
Phys. Lett. B \textbf{293}, 344 (1992)
\bibitem{5}
S. Meljanac, Z. Skoda, M. Stojic, arXiv:math.QA/1409.8188
(2014)
\bibitem{6}
S. Meljanac, Z. Skoda, D. Svrtan, SIGMA \textbf{8}, 013 (2012)
\bibitem{7}
S. Kresic-Juric, S. Meljanac, M. Stojic, Eur. Phys. J. C
\textbf{51}, 229 (2007)
\bibitem{8}
P. Vitale, J. Christophe Wallet, JHEP \textbf{04},
115 (2013)
\bibitem{9}
A. Gere, T. Juric, J. Christophe Wallet,
JHEP \textbf{12}, 045 (2015).
\bibitem{Ana2}
A. Pacho{\l}, arXiv:math-ph/1112.5366 (2011)
\bibitem{10}
T. Juric, S. Meljanac, A. Samsarov,
J. Phys. Conf. Ser. 634 (2015) 1, 012005
\bibitem{11}
Paolo Aschieri, arXiv:hep-th/0703013 (2007)
\bibitem{12}
P. Aschieri, M. Dimitrijevic, P. Kulish, F. Lizzi, J. Wess,
``Noncommutative Spacetimes'', Lecture Notes in Physics , Vol.
774 ISBN: 978-3-540-89792-7, Springer, (2009)
\bibitem{QFT1}
A. Agostini, G. Amelino-Camelia and F. D'Andrea,
Int. J. Mod. Phys. A \textbf{19}, 5187 (2004)
\bibitem{QFT2}
M. Dimitrijevic, L. Jonke, L. Moeller,
E. Tsouchnika, J. Wess and M. Wohlgenannt,
Eur. Phys. J. C \textbf{31}, 129 (2003)
\bibitem{QFT3}
G. Amelino-Camelia and M. Arzano, Phys. Rev.
D \textbf{65}, 084044 (2002)
\bibitem{QFT4}
P. Kosinski, J. Lukierski, P. Maslanka,
Phys. Rev. D \textbf{62}, 025004 (2000)
\bibitem{QFT5}
J. Lukierski, H. Ruegg and W. J. Zakrzewski
Annals. Phys. \textbf{243}, 90 (1995)
\bibitem{QFT6}
S. Meljanac, A. Samsarov, J. Trampetic, M. Wohlgenannt,
JHEP \textbf{1112}, 010 (2011)
\bibitem{QFT7}
S. Meljanac, A. Samsarov, Int. J. Mod. Phys. A \textbf{26},
1439 (2011)
\bibitem{QFT8}
H. Grosse, M. Wohlgenannt, Nucl. Phys. B \textbf{748}
, 473 (2006)
\bibitem{QFT9}
L. Freidel, J. Kowalski-Glikman, S. Nowak,
Phys. Lett. B \textbf{648}, 70 (2007)
\bibitem{QFT10}
J. Kowalski-Glikman, A. Walkus,
Mod. Phys. Lett. A \textbf{24}, 2243 (2009)
\bibitem{G}
G. Amelino-Camelia, Phys. Lett. B \textbf{510},
255 (2001)\\
G. Amelino-Camelia, Int. J. Mod. Phys. D \textbf{11},
35 (2002)
\bibitem{Smo}
J. Magueijo and L. Smolin, Phys. Rev. Lett.
\textbf{88}, 190403 (2002); J. Magueijo and L. Smolin,
Phys. Rev. D \textbf{67}, 044017 (2003).
\bibitem{J1}
J. Kowalski-Glikman and S. Nowak, Int. J. Mod.
Phys. D \textbf{12} 299 (2003)
\bibitem{Ghosh}
S. Das, S. Ghosh, D. Roychowdhury, Phys. Rev.
D \textbf{80}, 125036 (2009)\\
S. Das, D. Roychowdhury Phys.
Rev. D \textbf{81}, 085039 (2010)
\bibitem{Pai}
G. Amelino-Camelia and T. Pairan, Phys.
Rev. D \textbf{64}, 036005 (2001)
\bibitem{B1}
S. Corley, Phys. Rev. D \textbf{57}, 6280 (1998)
\bibitem{B2}
S. Corley and T. Jacobson, Phys. Rev. D \textbf{59},
124011 (1999)
\bibitem{B3}
A. Blaut, J. Kowalski-Glikman, D. Nowak-Szczepaniak,
Phys. Lett. B \textbf{521}, 364 (2001).
\bibitem{p1}
G. Salesi, Phys. Rev. D \textbf{85}, 063502 (2012)
\bibitem{p2}
J. Kowalski-Glikman arXiv:hep-th/0209264 (2002)
\bibitem{d1}
S. Sarkar, Mod. Phys. Lett. A \textbf{17}, 1025 (2002)
\bibitem{d2}
R. Aloisio, P. Blasi, A. Galante, P. L. Ghia, A.
F. Grillo, Astropart. Phys. \textbf{19}, 127 (2003)
\bibitem{d3}
A. Borowiec, Kumar S. Gupta, S. Meljanac, A. Pachol,
Europhys. Lett. \textbf{92}, 20006 (2010)
\bibitem{L1}
G. Amelino-Camelia, Nature \textbf{418}, 34 (2002)
\bibitem{L2}
J. Kowalski-Glikman,
Phys. Lett. B \textbf{547}, 291 (2002)
\bibitem{L3}
N. R. Bruno, G. Amelino-Camelia, J. Kowalski-Glikman,
Phys. Lett. B \textbf{522}, 133, (2001)
\bibitem{L4}
J. Kowalski-Glikman, Phys. Lett. A \textbf{286}, 391 (2001)
\bibitem{re1}
G. Amelino-Camelia, D. Benedetti, F. D'Andrea,
A. Procaccini. Class. Quant. Grav. \textbf{20}, 5353 (2003)
\bibitem{re2}
J. Kowalski-Glikman, arXiv:hep-th/0209264 (2002)
\bibitem{re3}
J. Kowalski-Glikman, Lect. Notes Phys.
\textbf{669}, 131 (2005)
\bibitem{re4}
G. Amelino-Camelia, J. Kowalski-Glikman,
G. Mandanici, A. Procaccini, Int. J. Mod.
Phys. A \textbf{20}, 6007 (2005)
\bibitem{Fermi}
A. Abdo et al., Science \textbf{323}, 1688 (2009)
\bibitem{2009}
G. Amelino-Camelia and L. Smolin, Phys.
Rev. D \textbf{80}, 084017 (2009)
\bibitem{al1}
P. Kosinski, J. Lukierski, P. Maslanka,
J. Sobczyk, Mod. Phys. Lett. A \textbf{10}, 2599
(1995)
\bibitem{al2}
J. Kowalski-Glikman, S. Nowak, Phys.
Lett. B \textbf{539}, 126 (2002)
\bibitem{al3}
J. Lukierski and A. Nowicki, Acta Phys.
Polon. B \textbf{33}, 2537 (2002)
\bibitem{al4}
S. Mignemi, Phys. Lett. B \textbf{672}, 186
(2009)
\bibitem{De1}
S. Judes, M. Visser, Phys. Rev. D
\textbf{68}, 045001 (2003)
\bibitem{De2}
S. Ghosh, P. Pal, Phys. Rev. D \textbf{75},
105021 (2007)
\bibitem{De3}
S. Meljanac, A. Samsarov, M. Stojic, K. S. Gupta
Eur. Phys. J. C \textbf{53}, 295 (2008)
\bibitem{De4}
T. R. Govindarajan, Kumar S. Gupta, E. Harikumar,
S. Meljanac, D. Meljanac
Phys. Rev. D \textbf{77} (2008) 105010
\bibitem{De5}
T. Juric, S. Meljanac, R. Strajn
Int. J. Mod. Phys. A \textbf{29}, 1450022 (2014)
\bibitem{De6}
S. Meljanac, A. Pachol, A. Samsarov, Kumar. S. Gupta
Phys. Rev. D \textbf{87},  125009 (2013)
\bibitem{De7}
S. Meljanac, M. Stojic
Eur. Phys. J. C \textbf{47}, 531 (2006)
\bibitem{sh}
N. Jafari, A. Shariati, AIP Conf.
Proc. \textbf{841}, 462 (2006)
\bibitem{Am}
G. Amelino-Camelia, Symmetry \textbf{2}, 230
(2010)
\bibitem{sig}
S. Das, S. Pramanik, S. Ghosh, SIGMA
\textbf{10}, 104 (2014)
\bibitem{GRB}
G. Amelino-Camelia and S. Majid, Int.
J. Mod. Phys. A \textbf{15}, 4301 (2000)
\bibitem{Text}
Keith R. Symon, ``Mechanics", ISBN: 978-0201073928
, 3rd ed., Addison-Wesley, (1971)\\
S. T. Thornton and J. B. Marion, ``Classical Dynamics of Particles and Systems",
ISBN: 964-01-0783-2, 3rd ed., Saunders College Publishing/Harcourt Brace, (1988).
\bibitem{Bala}
A. P. Balachandran, A. G. Martins and P. Teotonio-Sobrinho, JHEP \textbf{0705}, 066 (2007)
\bibitem{Tnx} This part of the text is taken directly from the referees report with minor changes.
\end{thebibliography}
\end{document}